\documentclass{ifacconf}

\usepackage{graphicx}      %
\usepackage{natbib}        %
\usepackage{amsmath}
\usepackage{amssymb}
\usepackage{bm}
\usepackage{mathtools}
\usepackage{siunitx}
\usepackage{glossaries}

\usepackage{algorithm}
\usepackage{algpseudocode}
\usepackage{lipsum}

\usepackage{subcaption}
\usepackage{subfig}

\usepackage[short,nocomma]{optidef}

\usepackage{tikz}
\usepackage{circuitikz}
\usepackage{graphicx}
\usetikzlibrary{positioning} 
\usetikzlibrary{fit}
\usetikzlibrary{calc}
\usetikzlibrary{spy}
\usetikzlibrary{plotmarks}
\usetikzlibrary{arrows.meta}
\usetikzlibrary{shapes.misc}
\usepackage{pgfplots}
\pgfplotsset{compat=newest}
\usepgfplotslibrary{patchplots}

\usepackage{pdfpages}

\begin{document}
\begin{frontmatter}

\title{Stability-informed Bayesian Optimization for MPC Cost Function Learning\thanksref{footnoteinfo}}
\thanks[footnoteinfo]{© 2024 the authors. This work has been accepted to NMPC 2024, IFAC for publication under a Creative Commons Licence CC-BY- NC-ND.}

\author[First]{Sebastian Hirt}
\author[First]{Maik Pfefferkorn}
\author[Third]{Ali Mesbah}
\author[First]{Rolf Findeisen} 

\address[First]{Control and Cyber-Physical Systems Laboratory, \\ Technical University of Darmstadt, Darmstadt, Germany, \\ \{sebastian.hirt, maik.pfefferkorn, rolf.findeisen\}@iat.tu-darmstadt.de.}
\address[Third]{Department of Chemical and Biomolecular Engineering, University of California, Berkeley, CA 94720, USA, mesbah@berkeley.edu.} 

\begin{abstract}
Designing predictive controllers towards optimal closed-loop performance while maintaining safety and stability is challenging.
This work explores closed-loop learning for predictive control parameters under imperfect information while considering closed-loop stability. 
We employ constrained Bayesian optimization to learn a model predictive controller's (MPC) cost function parametrized as a feedforward neural network, optimizing closed-loop behavior as well as minimizing model-plant mismatch.
Doing so offers a high degree of freedom and, thus, the opportunity for efficient and global optimization towards the desired and optimal closed-loop behavior. 
We extend this framework by stability constraints on the learned controller parameters, exploiting the optimal value function of the underlying MPC as a Lyapunov candidate.
The effectiveness of the proposed approach is underlined in simulations, highlighting its performance and safety capabilities.
 
\end{abstract}

\begin{keyword}
Closed-loop Learning, Model Predictive Control, Bayesian Optimization, Lyapunov Stability, Policy Optimization, Neural Network
\end{keyword}

\end{frontmatter} 

\section{Introduction}
\vspace{-5mm}
Model predictive control (MPC) is a widely employed control method that allows for optimal control of nonlinear systems, while explicitly considering constraints on the states and the inputs. Rooted in iterative prediction and optimization of the future system behavior, MPC involves solving a finite-horizon optimal control problem at each sampling instant \citep{rawlings2017model, findeisen2002introduction}. Thus, for stable and robust operation of a controlled plant, an accurate model is necessary to provide suitable predictions of the future system behavior. However, accurate models are not always available due to complex dynamics or lacking process insights. To overcome this challenge, many works explore machine learning for MPC, e.g., \citep{hewing2020cautious, maiworm2021online}, to learn an accurate, data-based model of the underlying plant. Doing so, however, introduces challenges with respect to stability and repeated feasibility, as MPC must be formulated robustly against model uncertainties, which is often costly and leads to overly conservative closed-loop behaviors. Furthermore, the accuracy of the prediction model, see, e.g., \citep{gevers1993towards, kordabad2023reinforcement}, does not necessarily translate to optimal closed-loop performance, e.g., reflected by a superordinate cost function.
In response to these challenges, recent works have investigated hierarchical learning and tuning frameworks for optimization-based controllers. These frameworks typically feature an upper-optimization layer for overall closed-loop performance optimization exploiting algorithms from reinforcement learning \citep{kordabad2023reinforcement, zanon2021safe, anand2023painless},  or Bayesian optimization \citep{paulson2023tutorial, piga2019performance, makrygiorgos2022performanceorienteda}. Simultaneously, a lower-level controller is responsible for short-term planning and control, and can, e.g., be realized by MPC with a limited prediction horizon. Such a hierarchical structure allows to achieve the desired closed-loop behavior, mitigating potential performance drawbacks due to inadequate prediction models or unnecessarily long prediction horizons to match local and global performance.

This work explores using prediction models that may show significant model-plant mismatch and MPC with short prediction horizon not spanning the complete time of an episode, such as a set-point change that should be optimized.
The long-term closed-loop behavior of the considered nonlinear dynamical system is defined by the performance over a whole episode. Specifically, Bayesian optimization, a global, sample-efficient optimization approach, is used to learn the stage cost of an MPC such that a desired long-term performance is optimized.
The MPC stage cost is parametrized by a feedforward artificial neural network, leveraging the expansive representation capabilities of neural networks for a broad class of functions. Learning the cost function provides an alternative approach to model learning, as model-based predictions are ultimately reflected in the cost function. Related ideas have been proposed in \citep{seel2022convex}, where reinforcement learning was employed. The parameter space in \citep{seel2022convex} is constrained to input convex neural networks and lacks stability certificates for the closed loop, motivating our contribution of a Lyapunov-informed, Bayesian optimization-based parameter learning procedure.

For learning the MPC stage cost, our procedure does not only take the closed-loop performance into account.
It furthermore provides a Lyapunov stability certificate for the resulting closed-loop system.
To this end, we incorporate stability information observed in the closed-loop into the Bayesian optimization layer via soft constraints.
Doing so, allows to guide the optimization towards safe and performant closed-loop behavior, ultimately yielding a stable controller. Specifically, we exploit the optimal cost of the MPC as a Lyapunov function candidate, as for example done in \citep{chang2019neural, berkenkamp2016safeb}. However, to the best of our knowledge, this has not been applied in the context of closed-loop learning for MPC.

The remainder of this paper is structured as follows.
We provide an overview of model predictive control, Gaussian process regression and Bayesian optimization in Section \ref{sec:fundamentals}.
Thereafter, the proposed approach is introduced in Section \ref{sec:main}, including the neural-network-based parameterization of the MPC scheme and the incorporation of stability constraints at the Bayesian optimization level.
In Section \ref{sec:simulation}, we underline the effectiveness of the proposed approach using simulations before drawing conclusions in Section \ref{sec:conclusion}.

\vspace{-1mm}
\section{Problem Formulation}
\vspace{-3mm}
\label{sec:fundamentals}
In this section, we outline the control task.
Afterwards, we introduce parametrized model predictive control, followed by an overview of Gaussian process regression.
We conclude by presenting the basics of constrained Bayesian optimization.

\vspace{-1mm}
\subsection{Control Task}
\vspace{-2mm}
We consider nonlinear, discrete-time dynamical systems
\begin{equation} 
\label{eqn:discrete_system_general}
    x_{k+1} = f(x_k, u_k),
\end{equation}
where $x_k \in \mathbb{R}^{n_\text{x}}$ are the system states, $u_k \in \mathbb{R}^{n_\text{u}}$ are the system inputs, $f: \mathbb{R}^{n_\text{x}} \times \mathbb{R}^{n_\text{u}} \to \mathbb{R}^{n_\text{x}}$ describes the dynamics and $k \in \mathbb{N}_0$ is the discrete time index.
The control objective is (i) to optimally steer system \eqref{eqn:discrete_system_general} from an initial set-point $(x_0, u_0)$ to a new reference set-point $(x_d, u_d)$ while satisfying input and state constraints, and (ii) to stabilize the system there.
To this end, we employ model predictive control.
We assume that the dynamics \eqref{eqn:discrete_system_general} are not exactly known, such that there is a model-plant mismatch between the true system and the prediction model used in the MPC.
To account for this mismatch, we learn the cost function of the MPC scheme, thereby introducing degrees of freedom for optimization of the long-term closed-loop performance.
The objective is then to select the parameterization of the cost function that leads to the best closed-loop performance as well as closed-loop stability, despite the model-plant mismatch.
To this end, we employ Bayesian optimization and infer the optimal parameterization from closed-loop data.
Stability certificates are obtained by imposing Lyapunov-like constraints on the Bayesian optimization cost function.

\vspace{-1mm}
\subsection{Parametrized Model Predictive Control}
\vspace{-2mm}
We consider an MPC formulation with $n_{\text{p}} \in \mathbb{N}$ learnable parameters $\theta \in \Theta \subset \mathbb{R}^{n_{\text{p}}}$.
This specifically includes, but is not restricted to, the parametrization of the cost function.
The MPC solves at every time index $k$ the parameterized optimal control problem
\begin{mini!}
    {\mathbf{\hat{u}}_k}{\left\{ J(x_k, \mathbf{\hat{u}}_k) = \sum_{i=0}^{N-1} l_\theta({\hat x}_{i \mid k}, {\hat u}_{i \mid k}) + E_\theta({\hat x}_{N \mid k}) \! \right\}\label{eqn:mpc_ocp_cost}}{\label{eqn:mpc_ocp}}{}
    \addConstraint{\forall i}{\in \{0, 1, \dots, N-1\}: \notag}{}
    \addConstraint{}{\hat x_{i+1\mid k} = \hat f_\theta(\hat x_{i\mid k}, \hat u_{i\mid k}), \ \hat x_{0 \mid k} = x_k,}{\label{eqn:mpc_ocp_model}}
    \addConstraint{}{\hat x_{i \mid k} \in \mathcal{X}_\theta, \ {\hat u}_{i \mid k} \in \mathcal{U}_\theta, \ \hat x_{N \mid k} \in \mathcal{E}_\theta.}{\label{eqn:mpc_ocp_constraints}}
\end{mini!}
Herein, $\hat{\cdot}_{i\mid k}$ denotes the model-based $i$-step ahead prediction at time index $k$, $\hat f_\theta(\cdot)$ is the (parametrized) prediction model and $x_k$ is the measurement of the system state at time index $k$.
Furthermore, the length of the prediction horizon is $N < \infty$ and $l_\theta(\cdot)$ and $E_\theta(\cdot)$ are the (parametrized) stage and terminal cost functions, respectively.
The constraints \eqref{eqn:mpc_ocp_constraints} are comprised of the (parametrized) state, input, and terminal sets $\mathcal{X}_\theta$, $\mathcal{U}_\theta$, and $\mathcal{E}_\theta$, respectively.
Solving \eqref{eqn:mpc_ocp} yields the optimal input sequence $\mathbf{\hat{u}}_k^*=[\hat u_{0 \mid k}^*,\dots,\hat u_{N-1 \mid k}^*]$, of which the first element is applied to system \eqref{eqn:discrete_system_general}. 
This way, repeatedly computing \eqref{eqn:mpc_ocp} at any time index $k$ defines the parameterized control policy $\hat u_{0 \mid k}^*$, which implicitly depends on the parameters $\theta$.

\vspace{-1mm}
\subsection{Gaussian Process Surrogate Models}
\vspace{-2mm}
Bayesian optimization is used to find optimal MPC parameters from closed-loop data.
We use Gaussian process surrogate models in the Bayesian optimization to learn an approximation of the not explicitly known closed-loop performance cost with respect to the parameters $\theta$.

Gaussian process (GP) regression enables to learn a probabilistic surrogate model of an unknown function $\varphi: \mathbb{R}^{n_\xi} \to \mathbb{R}, \xi \mapsto \varphi(\xi)$.
Loosely speaking, a GP $g(\xi) \sim \mathcal{GP}(m(\xi), k(\xi, \xi^\prime))$ defines a Gaussian probability distribution over functions.
Hence, sampling a GP yields functions, each of which is a possible realization of the underlying, unknown function $\varphi$.
The GP model is fully defined by its prior mean function $m: \mathbb{R}^{n_\xi} \to \mathbb{R}, \xi \mapsto \mathrm{E}[g(\xi)]$ and prior covariance function $k: \mathbb{R}^{n_\xi} \times \mathbb{R}^{n_\xi} \to \mathbb{R}, (\xi, \xi') \mapsto \mathrm{Cov}[g(\xi), g(\xi')]$.

The objective is to predict an unobserved function value $\varphi(\xi_*)$, the test target, at a test input $\xi_*$.
To obtain a meaningful prediction model, we rely on a set of (noisy) function observations $\mathcal{D} = \{ (\xi_i, \gamma_i = \varphi(\xi_i) + \varepsilon_i) \mid i \in \{1, \dots, n_{\text{d}} \}, \varepsilon_i \sim \mathcal{N}(0, \sigma^2) \}$.
Therein, $\varepsilon_i$ models white Gaussian noise with variance $\sigma^2$.
For notational convenience, we collect all training inputs $\xi_i$ in the input matrix $\Xi \in \mathbb{R}^{n_{\text{d}} \times n_\xi}$ and the corresponding training targets in the vector $\gamma \in \mathbb{R}^{n_{\text{d}} \times 1}$ and write equivalently $\mathcal{D} = (\Xi, \gamma)$.
Through Bayesian inference, the information provided by these so-called training data is incorporated in the model.

This inference step yields the posterior GP model $\varphi(\xi_*) \mid \Xi, \gamma, \xi_* \sim \mathcal{N}(m^+(\xi_*), k(\xi_*, \xi_*))$ with
\begin{subequations}
\begin{align}
    m^+(\xi_*) &= m(\xi_*)+k(\xi_*,  \Xi) k_{\gamma}^{-1} (\gamma-m(\Xi)), \label{eq:gp_postMean} \\
    k^+(\xi_*) &= k(\xi_*, \xi_*) - k(\xi_*, \Xi) k_{\gamma}^{-1} k(\Xi, \xi_*), \label{eq:gp_postVar}
\end{align}
\end{subequations}
where $k_{\gamma} = k(\Xi, \Xi) + \sigma^2 I$ and $I$ denotes the identity matrix.
The posterior mean \eqref{eq:gp_postMean} is an estimate for the unknown function value $\varphi(\xi_*)$, while the posterior variance \eqref{eq:gp_postVar} quantifies the prediction uncertainty \citep{rasmussen2006gaussian}.

The prior mean and covariance function are design choices and often depend on free hyperparameters, which have to be carefully adapted to the underlying problem.
One way of obtaining suitable hyperparameters is to infer them from the training data via evidence maximization, see \citep{rasmussen2006gaussian}.

\vspace{-1mm}
\subsection{Constrained Bayesian Optimization}
\vspace{-2mm}
Bayesian optimization (BO) is a sample efficient, iterative optimization scheme suited for optimization of black-box functions that are costly to evaluate.
In this work, BO is used to maximize the closed-loop performance subject to constraints as a function of the free parameters $\theta$ of the MPC \eqref{eqn:mpc_ocp}, which cannot be expressed in closed-form and is costly to evaluate.

Formally, the objective is to determine the optimal parameters $\theta^*$ by solving
\begin{align*}
    \theta^* &= \arg \max_{\theta \in \Theta} \left\{G_0(\theta)\right\} \\
    & \text{s.t.} \ \forall i \in \{1, \ldots, n_{\text{bc}} \}: G_i(\theta) \geq 0,
\end{align*}
where $G_0(\cdot)$ is the closed-loop performance and $G_i(\cdot), i \in \{1, \ldots, n_{\text{bc}} \}$ are $n_{\text{bc}} \in \mathbb{N}$ black-box constraints on the closed-loop realizations.
The lack of knowledge about the objective function $G_0$ or the constraints $G_i, i \in \{1, \ldots, n_{\text{bc}} \}$ prevents us from deploying standard optimization approaches such as gradient-based methods \citep{garnett2023bayesian}.
To overcome this situation, Bayesian optimization relies on Bayesian surrogate models of the unknown functions.
To this end, Gaussian process models are commonly used because of their effectiveness with a limited number of training data points, which aligns with our objective of a sample-efficient learning procedure.
These models are sequentially updated using samples of the unknown functions obtained for a series of query points.
Specifically, in each iteration $n \in \mathbb{N}_0$ of the Bayesian optimization, we
\begin{enumerate}
    \item[(i)] select a query point $\theta_n$ and conduct a closed-loop experiment using MPC \eqref{eqn:mpc_ocp} to generate a new data point $\{ \theta_n, G_0(\theta_n), \ldots, G_{n_{\text{bc}}}(\theta_n) \}$,
    \item[(ii)] augment the training data set by the new data point according to $\mathcal{D}_{n+1} \leftarrow \mathcal{D}_n \cup \{ \theta_n, G_0(\theta_n), \ldots, G_{n_{\text{bc}}}(\theta_n) \}$, and
    \item[(iii)] update the (posterior) GP models using $\mathcal{D}_{n+1}$, including hyperparameter optimization.
\end{enumerate}
Herein, subscript $n$ denotes the dependency on the current iteration and $\mathcal{D}_n$ is the set of training data collected up to iteration $n$.

To effectively guide the sequential search, i.e., the selection of new query points $\theta_{n}$, towards the optimizer $\theta^*$ with increasing $n$, we rely on acquisition functions.
An acquisition function $\alpha: \mathbb{R}^{n_{\text{p}}} \to \mathbb{R}, \theta \mapsto \alpha(\theta; \mathcal{D}_n)$ exploits the surrogate models of the unknown functions to evaluate the utility of a query point $\theta_n$ based on the available knowledge up to iteration $n$, trading off exploration and exploitation in the parameter space.
The next query point is then determined according to
\begin{equation}
    \label{eqn:bo_update}
    \theta_{n+1} = \arg \max_{\theta \in \Theta} \left\{ \alpha(\theta; \mathcal{D}_n) + \epsilon \ \beta(\theta; \mathcal{D}_n) \right\}.
\end{equation}
Here, $\beta: \mathbb{R}^{n_{\text{p}}} \to \mathbb{R}, \theta \mapsto \beta(\theta; \mathcal{D}_n)$ is a penalty term that is used to incorporate the black-box constraints $G_i(\cdot), i \in \{1, \ldots, n_{\text{bc}} \}$ into the BO procedure and $\epsilon \in \mathbb{R}_+$ is a penalty parameter.
The penalty term can be defined in different ways, e.g., as a log-barrier term \citep{krishnamoorthy2023tuning} or as a soft constraint.

\vspace{-1mm}
\section{Lyapunov-informed Learning of a Neural Cost Function}
\vspace{-3mm}
\label{sec:main}
In this section, we first outline our approach of using a neural stage cost function and explain how its parameters are learned via Bayesian optimization.
This is followed by a discussion of how we exploit information about Lyapunov stability during the learning procedure, yielding a stable and safe closed-loop system.

\vspace{-1mm}
\subsection{Neural Cost Function}
\vspace{-2mm}
In practice, accurate prediction models $\hat f(x, u)$ for the system dynamics $f(x, u)$ are not always available.
Thus, we refrain from the usual approach of learning an open-loop dynamics model from data.
Instead, we employ a more direct approach and learn the MPC stage cost directly from closed-loop experiments.
This is motivated by the idea that the prediction model is ultimately reflected in the cost function according to $J(x_k, \mathbf{\hat{u}}_k) = \sum_{i=0}^{N-1} l({\hat x}_{i \mid k}, {\hat u}_{i \mid k}) + E({\hat x}_{N \mid k}) =  l(x_k, \hat{u}_{0 \mid k}) + \sum_{i=1}^{N-1} l \big( \hat{f}(\hat{x}_{i-1 \mid k}, \hat{u}_{i-1 \mid k}), \hat{u}_{i \mid k} \big) + E \big( \hat{f}(\hat{x}_{N-1 \mid k}, \hat{u}_{N-1 \mid k}) \big)$.
We employ a quadratic stage cost function for set-point regulation
\begin{equation}
    l(x,u) = (x-x_d)^\top Q (x-x_d) + (u-u_d)^\top R (u-u_d)
\end{equation}
with fixed weighting matrices $Q \in \mathbb{R}^{n_{\text{x}} \times n_{\text{x}}}, Q \succeq 0$ and $R \in \mathbb{R}^{n_{\text{u}} \times n_{\text{u}}}, R \succ 0$.
To account for model imperfections, we define the parameterized stage cost
\begin{equation}
\label{eqn:specific_stage_cost}
    l_\theta(x, u) = l(x,u) + l_{NN}(x; \theta).
\end{equation}
The term $l_{NN}(\cdot; \theta)$ is given by the output of a feedforward neural network with $L$ layers
\begin{equation*}
    y_{NN}(z) = W_L \sigma(W_{L-1} \sigma(...\sigma(W_1 z + b_1)... ) + b_{L-1}) + b_L.
\end{equation*}
Here, $y_{NN}: \mathbb{R}^{n_{\text{x}}} \to \mathbb{R}$, $\sigma: \mathbb{R} \to \mathbb{R}$ is the activation function and $W_i$ and $b_i$, $i = 1, \ldots, L$, are the weight matrix and bias vector of the transition from layer $i-1$ to layer $i$ (layer zero denotes the input layer).
Specifically, we define
\begin{equation}\label{eq:NN_cost_term}
    l_{NN}(x; \theta) = y_{NN}(x)-y_{NN}(x_d)
\end{equation}
to ensure that $l_\theta(x_d, u_d) = 0$.
We employ a quadratic terminal cost $E(\hat{x}_{N \mid k}) = (\hat{x}_{N \mid k} - x_d)^\top P (\hat{x}_{N \mid k} - x_d)$ based on the solution to the discrete-time Riccati equation $P \in \mathbb{R}^{n_x \times n_x}$ for the problem set-up and the linearized dynamics around $x_d$.
Given that me must have knowledge of the system in a neighborhood of $x_d$, we do not modify the terminal cost and use $E_\theta(\hat{x}_{N \mid k}) = E( \hat{x}_{N \mid k})$.

Consequently, the learnable parameters that may be adapted by the BO algorithm according to a closed-loop performance measure $G_0(\theta)$ are given by $\theta = \left\{ W_1, b_1, \dots, W_L, b_L \right\}$.
Note that this approach results in a high number of parameters, introducing a high-dimensional parameter space that has to be explored during learning.
This restricts the size of the FNN and a trade-off between computational complexity and expressiveness has to be found.
Note that the approach can be extended by additionally parameterizing the model or constraints.
Furthermore, satisfaction of the state constraints can not be guaranteed by the MPC due to the model-plant mismatch.
A detailed treatment is left out here for simplicity of presentation.
However, state constraints can be enforced probabilistically by incorporating them in the BO problem via a barrier term, see, e.g., \citep{krishnamoorthy2023tuning}.

\vspace{-1mm}
\subsection{Stability Guarantees via Lyapunov Constraints}
\vspace{-2mm}
When using BO to learn MPC parameters, it is not straightforward to obtain stability certificates.
However, stability is a crucial requirement for any control system.
Thus, we incorporate information about the Lyapunov stability of system \eqref{eqn:discrete_system_general} under MPC \eqref{eqn:mpc_ocp} into the learning procedure.
As a Lyapunov function candidate, we exploit the optimal value function, given by
\begin{equation}
    J^*(x_k) = J(x_k, \mathbf{\hat{u}}_k^*).
\end{equation}
For $J^*(x_k)$ to be a valid Lyapunov function for system \eqref{eqn:discrete_system_general}, the parameters $\theta$ have to be learned such that the following three conditions are satisfied \citep{rawlings2017model, findeisen2002introduction}:
\begin{subequations}
\begin{align}
    & J^*(x_d) = 0, \label{eqn:lyapunov_conditions_zero} \\
    & J^*(x_k) > 0, \ x_k \neq x_d, \label{eqn:lyapunov_conditions_positive_definite} \\
    & J^*(x_{k+1})-J^*(x_{k}) < 0, \ x_{k+1}, x_k \neq x_d. \label{eqn:lyapunov_conditions_decreasing}
\end{align}
\end{subequations}
Condition \eqref{eqn:lyapunov_conditions_zero} is trivially fulfilled for the chosen stage and terminal costs given \eqref{eq:NN_cost_term}.
We propose to formulate the conditions \eqref{eqn:lyapunov_conditions_positive_definite} and \eqref{eqn:lyapunov_conditions_decreasing} as black-box constraints in the BO problem. That is, we individually learn GP surrogate models for each constraint.

To this end, we define the following constraints at the BO layer for a closed-loop run of length $M \in \mathbb{N}$ as
\begin{subequations}
\begin{align}
    & G_1(\theta) \! = \! \sum_{k=0}^M \min\{ 0, J^*(x_k) \} \label{eqn:lyapunov_conditions_positive_definite_bo_constraint}, \\
    & G_2(\theta) \! = \! \sum_{k=0}^{M-1} \min\{ 0, -\left( J^*(x_{k+1})-J^*(x_k) \right) \} \label{eqn:lyapunov_conditions_decreasing_bo_constraint},
\end{align}
\end{subequations}
where \eqref{eqn:lyapunov_conditions_positive_definite_bo_constraint} encodes the positive definiteness constraint corresponding to \eqref{eqn:lyapunov_conditions_positive_definite} and \eqref{eqn:lyapunov_conditions_decreasing_bo_constraint} corresponds to \eqref{eqn:lyapunov_conditions_decreasing}, which penalizes if the Lyapunov function is not decreasing along the closed-loop trajectory.
The GPs for the Lyapunov constraints are learned based on the closed-loop trajectories obtained from deploying the MPC with the current parameters $\theta_n$ and incorporated into the optimization procedure as soft constraints via the penalty term $\beta(\cdot)$, see \eqref{eqn:bo_update}.
Note that this approach allows for violations of the Lyapunov conditions during the learning procedure but guides the optimizer towards a set of parameters that lead to a stable closed-loop behavior.
Further, we only check the constraints for all points $x_k$ along the observed closed-loop trajectory.
Hence, information about Lyapunov stability is only extracted from so-far observed closed-loop trajectories.
By assuming a continuous functional relationship between the parameter space and the constraint values, stability information can be extrapolated to nearby points in the parameter space, as necessary in the BO approach.

\vspace{-1mm}
\section{Simulation Results}
\vspace{-3mm}
\label{sec:simulation}
We illustrate the effectiveness of the proposed approach in simulation.
After introducing the set-up, we show closed-loop results when the cost function is learned with and without imposing stability constraints.

\vspace{-1mm}
\subsection{Set-Up}
\vspace{-2mm}
We consider a double pendulum with time-continuous dynamics given by
\begin{subequations}
\begin{align}\label{eq:pendulum_continuous}
\begin{split}
    \ddot \psi_1 =& (m_2  l_1 \dot{\psi}_1^2 s_{21} c_{21} + m_2 g s_2 c_{21} + m_2  l_2 \dot{\psi}_2^2 s_{21} \\
    &- (m_1 \! + \! m_2) g s_1) / (m_1 \! + \! m_2)  l_1 - m_2  l_1 c_{21}^2 + u 
\end{split} \\
\begin{split}
    \ddot \psi_2 =& (-m_2  l_2 \dot{\psi}_2^2 s_{21} c_{21} \! + \! (m_1 \! + \! m_2) (g s_1 c_{21} \! - \! l_1 \dot{\psi}_1^2 s_{21} \\
    &- g s_2)) / ( l_2 /  l_1) (m_1 \! + \! m_2)  l_1 - m_2  l_1 c_{21}^2,
\end{split}
\end{align}
\end{subequations}
where $s_i = \sin(\psi_i)$, $c_i = \cos(\psi_i)$ for $i \in \{1,2\}$ and $s_{21} = \sin(\psi_2 - \psi_1)$, $c_{21} = \cos(\psi_2 - \psi_1)$.
The system states $x = (\psi_1, \psi_2, \dot \psi_1, \dot \psi_2)$ are the angles $\psi_1, \psi_2$ of the links and the corresponding angular velocities $\dot{\psi}_1, \dot{\psi}_2$.
The control input $u \in [-50, 50]$ describes the acceleration acting at the base of the first link.
The parameters $m_1 = \SI{1}{\kg}, m_2 = \SI{1}{\kg}$ and $l_1 = \SI{1}{\m}, l_2 = \SI{1}{\m}$ describe the masses and lengths of the links.
We obtain the time-discrete dynamics using one-step fourth-order Runge-Kutta discretization over a sampling period of $T_s = \SI{0.05}{\s}$.
The control objective is to bring the pendulum in the upright position $x_d = (\pi, \pi, 0, 0), u_d = 0$ and to stabilize it there, starting from an initial position $x_0$.

We measure the closed-loop performance for a trajectory of length $M$ by the total weighted deviation of inputs and states from the set-point $(x_d, u_d)$, given by
\begin{equation}
\label{eqn:closed_loop_performance}
\begin{split}
    G_0(\theta) \! = \! \sum_{k=0}^{M} \lVert x_k \! - \! x_d \rVert_V^2 \! + \! \lVert u_k \! - \! u_d \Vert_W^2 \! + \! \lVert x_M \! - \! x_d \rVert_Z^2
\end{split}
\end{equation}
with weighting matrices $V, Z \in \mathbb{R}^{n_\text{x} \times n_\text{x}}, V \succeq 0, Z \succeq 0$ and $W \in \mathbb{R}^{n_\text{u} \times n_\text{u}}, W \succeq 0$.
Note that \eqref{eqn:closed_loop_performance} is structurally similar to the MPC cost function in our specific simulation setup, c.f. Section 3.1.
However, \eqref{eqn:closed_loop_performance} is based on the entire trajectory and thus global and long-term-oriented, while the MPC cost function considers only the near future defined by horizon $N$ and is thus local and short-term-oriented.
In general, other high-level closed-loop objectives can be encoded in \eqref{eqn:closed_loop_performance}.
In our simulations, we employ CasADi to implement the MPC problem, including the neural cost function, and BoTorch for Bayesian optimization.

\vspace{-1mm}
\subsection{Cost Function Learning without Stability Information}
\begin{figure}[t]
    \begin{subfigure}{0.5\textwidth}
        \includegraphics{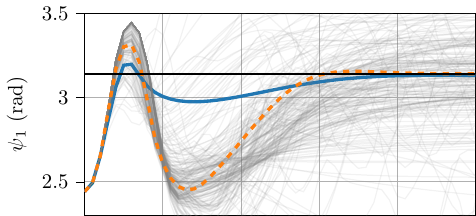}
    \end{subfigure}
    \hfill
    \begin{subfigure}{0.5\textwidth}
        \includegraphics{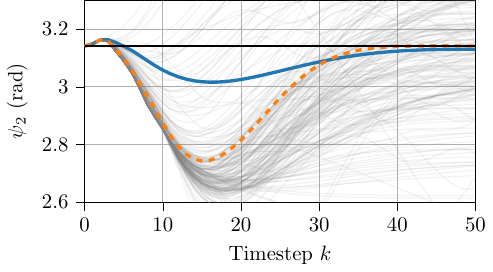}
        \vspace{-4mm}
    \end{subfigure}
    \caption{Nominal closed-loop state trajectory without cost function modification (orange dashed line), learned trajectory (blue solid line) and all other queried trajectories (gray solid lines) without enforcing stability constraints.}
    \label{fig:sim_unconstrained}
\end{figure}

\vspace{-2mm}
We consider learning the NN-parameterized MPC stage cost without imposing constraints \eqref{eqn:lyapunov_conditions_positive_definite_bo_constraint}, \eqref{eqn:lyapunov_conditions_decreasing_bo_constraint} as a basis of comparison, first.
In the MPC, we use a linear prediction model obtained from linearizing the nonlinear dynamics around $x_d$, thereby introducing a model-plant mismatch.
The stage cost is parameterized by a FNN with one hidden layer comprised of five neurons, resulting in $n_{\text{p}} = 31$ learnable parameters, and using the hyperbolic tangent activation function throughout the hidden layer.
The chosen network structure is sufficient to counteract the model-plant mismatch.
We compute an initial data set $\mathcal{D}_0$ using ten randomly sampled parameterizations and the corresponding closed-loop performance according to \eqref{eqn:closed_loop_performance} evaluated for the corresponding closed-loop trajectories.
Note that these trajectories are not necessarily stable and result in general in large deviations from the reference set-point.
We place a zero-mean GP prior with Matérn covariance function over $G_0(\theta)$ and derive suitable hyperparameters as well as the posterior model based on $\mathcal{D}_0$.
After initializing the required components, the BO-based learning procedure is run for 400 iterations using the expected improvement acquisition function.
In each iteration $n \in \{ 1, \ldots, 400 \}$, we compute the closed-loop trajectory for 50 time steps using the current parameterization $\theta_n$.
The state trajectories of all iterations are shown by gray lines in Figure \ref{fig:sim_unconstrained} (for the sake of brevity, we omit the results for $\dot{\psi}_1$ and $\dot{\psi}_2$).
Furthermore, the closed-loop-optimal state trajectory corresponding to the best found parameterization is highlighted in blue.
We compare this learned trajectory against the nominal trajectory that is obtained without the cost function modification, i.e., $\theta = 0$, and shown as orange dashed line in Figure \ref{fig:sim_unconstrained}.
Although both trajectories steer the system into the desired reference set-point and are hence stable, the reference set-point is reached earlier in the nominal case.
However, the learned trajectory shows fewer oscillations and overshooting and remains closer to the reference than the nominal trajectory, corresponding to the desired closed-loop behavior.
Note that although both trajectories appear to be stable, there are no formal stability certificates as we have neither enforced compliance with nor computed a Lyapunov function.
Furthermore, it is seen in Figure \ref{fig:sim_unconstrained} that many trajectories computed during the learning procedure diverge, i.e., are unstable, or end up in a different set-point than the desired one.
The former is explained by not enforcing stability constraints for the trajectories; the latter may be due to the MPC being stuck in local optima, as we do not enforce certain properties such as convexity of the modified cost function.

\vspace{-1mm}
\subsection{Lyapunov-informed Cost Function Learning}
\vspace{-2mm}
We now consider learning the NN-parameterized MPC cost function subject to the Lyapunov-informed stability constraints \eqref{eqn:lyapunov_conditions_positive_definite_bo_constraint} and \eqref{eqn:lyapunov_conditions_decreasing_bo_constraint}.
We employ the nonlinear dynamics with the parameter estimates $\hat{l}_1 = \hat{l}_2 = \SI{1.2}{\m}$, $\hat{m}_1 = \SI{2}{\kg}$ and $\hat{m}_2 = \SI{0.5}{\kg}$, introducing a model-plant mismatch due to the parameter error.
The stage cost is parameterized by a FNN with one hidden layer comprised of ten neurons, resulting in $n_p = 61$ learnable parameters, and using the hyperbolic tangent activation function throughout the hidden layer.
Note that the FNN is now of larger size than in the unconstrained case, which is to increase the expressiveness of the FNN in order to enable it to capture the increased complexity due to the constraints.

We place zero-mean Gaussian process priors with Matérn covariance functions over the performance measure $G_0(\theta)$ and the stability constraints $G_1(\theta), G_2(\theta)$.
Note that all GP models are continuously differentiable by the choice of the prior covariance function as we require the functional relationship between the parameters (the resulting state trajectories, respectively) and the Lyapunov constraints to be well-behaved.
However, the original constraint formulations \eqref{eqn:lyapunov_conditions_positive_definite_bo_constraint} and \eqref{eqn:lyapunov_conditions_decreasing_bo_constraint} possess nondifferentiable points due to taking a minimum.
Thus, the GP models of the constraints cannot exactly represent the boundaries of the true safe parameter set for which the constraints are satisfied.
To initialize the GP surrogate models, we use again an initial data set $\mathcal{D}_0$ comprised of ten randomly sampled parameterizations and the corresponding performance values according to \eqref{eqn:closed_loop_performance} and levels of constraint violations according to \eqref{eqn:lyapunov_conditions_positive_definite_bo_constraint} and \eqref{eqn:lyapunov_conditions_decreasing_bo_constraint}.
The BO-based learning procedure is run for 300 iterations with the expected improvement acquisition function.
The constraints are imposed as soft constraints according to \eqref{eqn:bo_update} with $\beta(\theta) = -\sum_{i=1}^2 \log(\exp(-1000 \ G_i(\theta))+1)$ and $\epsilon = 1$.

\begin{figure}[b]
    \centering
    \includegraphics{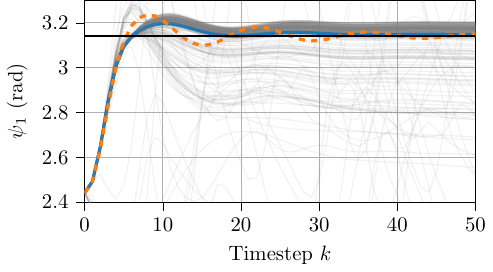}
    \vspace{-4mm}
    \caption{Nominal trajectory (orange dashed line), learned trajectory (blue solid line) and all other queried trajectories (gray solid lines) while imposing Lyapunov-like stability constraints.}
    \label{fig:sim_lyapunov}
\end{figure}

\begin{figure}[t]
    \centering
    \includegraphics{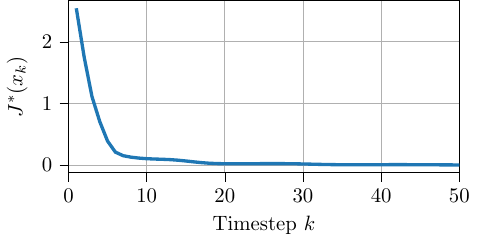}
    \vspace{-4mm}
    \caption{Evolution of the optimal value function along the optimal, learned closed-loop trajectory.}
    \label{fig:sim_lyapunov_lyapunov_function}
\end{figure}

The resulting closed-loop trajectories of all iterations are shown in Figure \ref{fig:sim_lyapunov} by gray lines.
Again, the nominal trajectory (orange dashed line) and the optimal, learned trajectory (blue solid line) are highlighted.
For the sake of brevity, we only show the results for $\psi_1$.
Both the nominal and the learned trajectory steer the system into the desired set-point.
However, the learned trajectory reaches the set-point faster and shows fewer oscillations and overshooting than the nominal trajectory.
Furthermore, the learned controller satisfies the Lyapunov stability constraints, i.e., the optimal value function of the MPC is a Lyapunov function along the considered trajectory (Figure \ref{fig:sim_lyapunov_lyapunov_function}).
Hence, a stability certificate for the considered scenario is provided.
The results indicate that the cost function modification can safely account for modeling errors and counteract them.

However, many of the trajectories that result from queried nonoptimal parameterizations -- especially the ones from early iterations -- violate the imposed stability constraints.
Although this seems counterintuitive at first sight, it is explained as follows.
Firstly, the randomly sampled initial data points do not necessarily include a parameterization that satisfies the constraints.
In consequence, the learning algorithm has no valid starting point to construct a series of safe parameterizations.
In fact, it is nontrivial to compute a safe initial parameterization for the proposed approach.
Secondly, there is in general no unique Lyapunov function for a system such that a trajectory might be stable, although the optimal value function is not a valid Lyapunov function for it.
Restricting ourselves to using the optimal value function as Lyapunov function introduces conservatism in the proposed approach that might be alleviated by additionally learning a suitable Lyapunov function.
Thirdly, the constraints are imposed as soft constraints, such that the learning algorithm is allowed to violate constraints if it expects a significant performance increase from doing so.

\vspace{-1mm}
\section{Conclusion}
\vspace{-3mm}
\label{sec:conclusion}
We have proposed automated learning of control policies, consisting of a Bayesian optimization layer and a lower-level parametrized model predictive controller. 
Specifically, Bayesian optimization was employed to systematically learn a parameterized cost function associated with the model predictive controller. 
The cost function was parameterized using a feedforward neural network, offering a flexible framework for learning the controller parameters such that a desired closed-loop behavior is achieved.
The latter may account for higher-level performance measures, as well as for compensation of model uncertainties that are often encountered in practice. 
We have proposed to directly account for stability via Lyapunov-informed constraints at the Bayesian optimization layer, ensuring satisfaction of the Lyapunov stability conditions along the resulting closed-loop trajectory by the learned controller.
The proposed approach was illustrated in simulation, showcasing both the performance and safety capabilities of the framework in a setting with significant model-plant mismatch. 

Future research will focus on establishing probabilistic stability and feasibility guarantees, exploiting algorithms from safe Bayesian optimization.
This will enable to probabilistically satisfy the stability constraints also during the learning procedure, i.e., for a large fraction of trajectories, when starting with an initial safe guess.
Finally, we also aim to examine closed-loop learning in higher-dimensional parameter spaces, with a focus on hybrid approaches of reinforcement learning and Bayesian optimization to exploit the advantages of both approaches.

\vspace{-4mm}
\bibliography{ifacconf}

\end{document}